\documentclass[twocolumn,showpacs,preprintnumbers,amsmath,amssymb]{revtex4-1}
\usepackage{graphicx}
\usepackage{dcolumn}
\usepackage{bm}
\usepackage{braket}
\usepackage[usenames,dvipsnames]{color}

\begin{document}

\title{Mott transition and magnetism of the triangular-lattice Hubbard model \\ with next-nearest-neighbor hopping}

\author{Kazuma Misumi, Tatsuya Kaneko, and Yukinori Ohta}
\affiliation{Department of Physics, Chiba University, Chiba 263-8522, Japan}

\date{\today}

\begin{abstract}
The variational cluster approximation is used to study the isotropic triangular-lattice 
Hubbard model at half filling, taking into account the nearest-neighbor ($t_1$) and 
next-nearest-neighbor ($t_2$) hopping parameters for magnetic frustrations.  We 
determine the ground-state phase diagram of the model.  In the strong correlation 
regime, the 120$^\circ$ N\'eel and stripe ordered phases appear, and a nonmagnetic 
insulating phase emerges in between.  In the intermediate correlation regime, 
the nonmagnetic insulating phase expands to a wider parameter region, which goes into 
a paramagnetic metallic phase in the weak correlation regime.  The critical phase 
boundary of the Mott metal-insulator transition is discussed in terms of the van Hove 
singularity evident in the calculated density of states and single-particle spectral function.  
\end{abstract}


\maketitle

\section{Introduction}

The physics of geometrical frustration in strongly correlated electron systems 
has long attracted much attention \cite{anderson,lee,balents}.  In particular, possible 
absence of magnetic long-range orders at zero temperature in the Heisenberg and 
Hubbard models defined on frustrated lattices, or the realization of a spin liquid phase 
as an exotic state of matter, has been one of the major issues in this field.  
The Mott metal-insulator transition is also a fundamental phenomenon in the field 
of strongly correlated electron systems \cite{mott,imada}, which has attracted much 
experimental and theoretical interest as well.  As one of the simplest models with 
geometrical frustration and Mott transition, we therefore study the Hubbard model 
at half filling defined on the triangular lattice in this paper, where not only the 
nearest-neighbor hopping parameters but also the next-nearest-neighbor ones are 
included.  

Much effort has so far been devoted in the study of the triangular-lattice Hubbard 
model with \textit{anisotropic} nearest-neighbor hopping parameters 
\cite{morita,sahebsara,watanabe,ohashi,tocchio1,tocchio2,yamada1,laubach,misumi}, which was 
motivated by experimental findings of possible spin liquid states in some organic Mott 
insulators such as $\kappa$-(ET)$_2$Cu$_2$(CN)$_3$ \cite{komatsu,shimizu,kurosaki,manna} 
and EtMe$_3$Sb[Pd(dmit)$_2$]$_2$ \cite{itou,yamashita}.  
The triangular-lattice Heisenberg model with the anisotropic exchange interactions 
has also been studied to find a variety of ordered phases such as N\'eel and spiral 
orders, as well as the quantum disordered (or spin liquid) phases in-between 
\cite{weihong,hauke1,hauke2}.  

However, to the best of our knowledge, the \textit{isotropic} triangular-lattice Hubbard 
model with both the nearest-neighbor ($t_1$) and next-nearest-neighbor ($t_2$) hopping 
parameters has not yet been addressed, the study of which will therefore provide useful 
information on the physics of magnetic frustrations and Mott metal-insulator transition 
in strongly correlated electron systems.  

The isotropic Heisenberg model with the nearest-neighbor ($J_1$) and next-nearest-neighbor 
($J_2$) exchange interactions, which may be derived by the second-order perturbation 
of the above-mentioned Hubbard model in the strong correlation limit, has on the other hand 
been studied much in detail, mostly from the theoretical point of view \cite{ma}.  
In the classical Heisenberg model where the spins are treated as classical vectors, 
it is known that a single phase transition occurs at $J_2/J_1=1/8$ between the three-sublattice 
120$^\circ$ N\'eel ordered state and an infinitely degenerate four-sublattice magnetically 
ordered states \cite{jolicoeur}.  This degeneracy is lifted by quantum fluctuations, thereby 
selecting a two-sublattice stripe ordered state through the so-called ``order-by-disorder'' 
mechanism \cite{jolicoeur,chubukov,deutscher,lecheminant}.  
One may then expect in the corresponding quantum Heisenberg model that an intermediate 
phase can appear near the classical critical point at $J_2/J_1=1/8$, for which many studies 
have been carried out to predict that the nonmagnetic disordered phase bordered by the 
120$^\circ$ N\'eel ordered phase at $J_2/J_1\simeq 0.05-0.12$ and the stripe ordered 
phase at $J_2/J_1\simeq 0.14-0.19$ actually emerges.  In particular, recent studies actually 
predict the emergence of either a gapless or gapped spin liquid phase in this intermediate 
region \cite{manuel,mishmash,kaneko,li2,zhu,hu2,iqbal}.  
These results of the Heisenberg model may be compared with those of our Hubbard model 
in the strong correlation limit (as we will see below).  

In this paper, motivated by the above developments in the field, we will study the triangular-lattice 
Hubbard model at half filling with the isotropic nearest-neighbor and next-nearest-neighbor 
hopping parameters in its entire interaction strength.  We use the variational cluster 
approximation (VCA), one of the quantum cluster methods based on the self-energy functional 
theory (SFT) \cite{potthoff1,potthoff2,dahnken,potthoff3,senechal}, which 
enables us to take into account the quantum fluctuations of the model with geometrically 
frustrated spin degrees of freedom.  We thereby calculate the grand potential of the system 
as a function of the Weiss fields for spontaneous symmetry breakings; here, we take the 
120$^\circ$ N\'eel and stripe magnetic orders and evaluate the order parameters and critical 
interaction strengths.  We also calculate the charge gap as well as the density of states 
(DOS) and single-particle spectral function using the cluster perturbation theory 
(CPT) \cite{senechal} and determine the ground-state phase diagram of the model in its 
entire parameter region.  

We will thereby show that in the strong correlation regime the 120$^\circ$ N\'eel and stripe 
ordered phases appear, and in-between, the nonmagnetic insulating phase caused by the 
quantum fluctuations in the frustrated spin degrees of freedom emerges, in agreement with 
the Heisenberg model studies.  We will also show that in the intermediate correlation regime 
the nonmagnetic insulating phase expands to wider parameter regions located around 
$0\le t_2/t_1\alt 0.3$ and $0.4\alt t_2/t_1\le 1$, which go into a paramagnetic metallic 
phase in the weak correlation regime via the second-order Mott transition.  
The characteristic behavior of the critical phase boundary of the Mott transition is discussed 
in terms of the van Hove singularity appearing in the calculated DOS and single-particle 
spectral function.  

The rest of the paper is organized as follows.  
In Sec.~II, we introduce the model and discuss the method of calculation briefly.  
In Sec.~III A, we present our results obtained in the strong correlation regime and 
compare them with those of the Heisenberg model.  
In Sec.~III B, we present our results obtained in the intermediate to weak correlation 
regime and discuss the phase diagram of our model.  The critical phase boundary of 
the Mott transition is also discussed.  A summary of the paper is given in Sec.~IV.  

\begin{figure}[thb]
\begin{center}
\includegraphics[width=0.9\columnwidth]{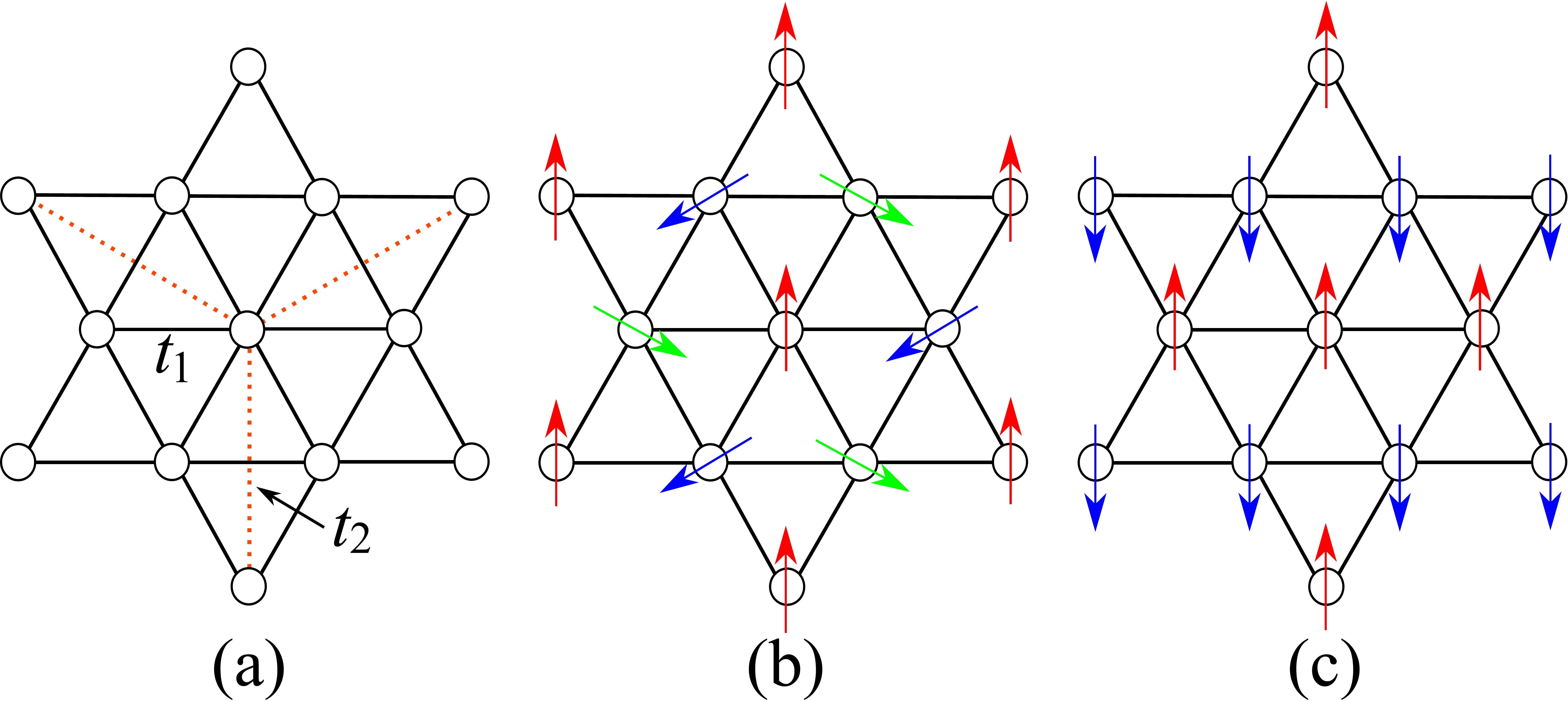}
\caption{(Color online) 
Schematic representations of 
(a) the triangular-lattice Hubbard model with the nearest-neighbor ($t_1$) and next-nearest-neighbor ($t_2$) hopping parameters, 
(b) the 120$^\circ$ N\'eel order, and 
(c) the stripe order.  
The arrows represent the directions of electron spins on the A, B, and C sublattices defined by different colors.
}\label{fig1}
\end{center}
\end{figure}

\begin{figure}[thb]
\begin{center}
\includegraphics[width=\columnwidth]{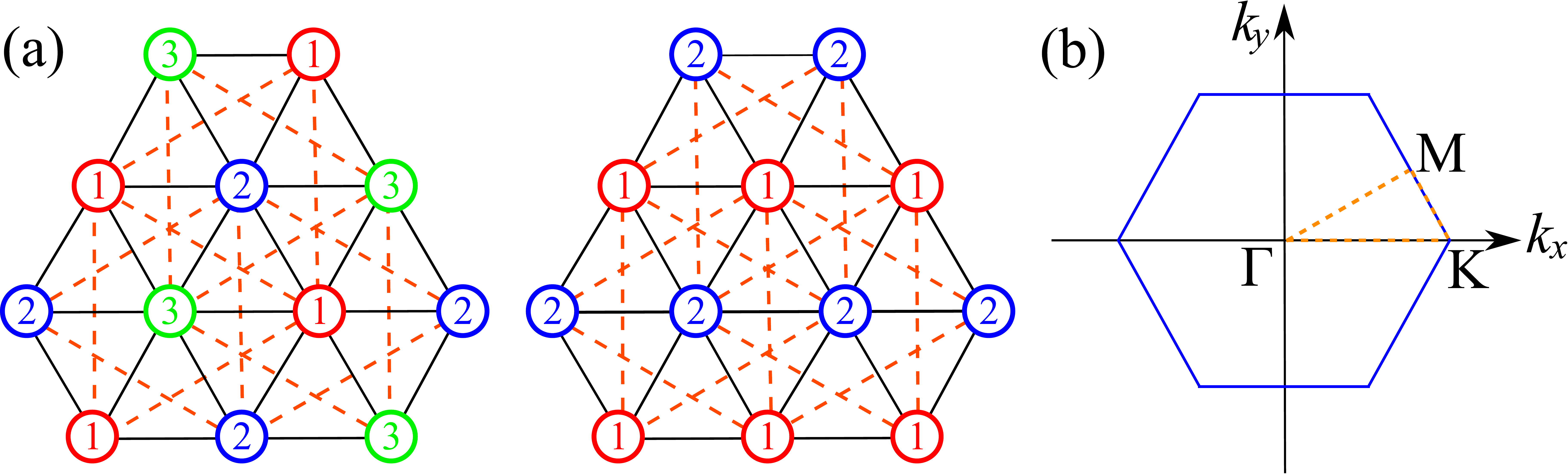}
\caption{(Color online) 
(a) The reference system of the 12-site cluster used in our analysis; 
the three-sublattice system corresponding to the 120$^\circ$ N\'eel order (left) 
and the two-sublattice system corresponding to the stripe order (right).  
(b) The first Brillouin zone of our triangular-lattice Hubbard model: 
$\Gamma(0,0)$, $\textrm{K}(4\pi/3,0)$, and $\textrm{M}(\pi,\pi/\sqrt{3})$.  
}\label{fig2}
\end{center}
\end{figure}

\section{Model and Method}

We consider the triangular-lattice Hubbard model [see Fig.~\ref{fig1}(a)] defined by the Hamiltonian 
\begin{align}
H=&-t_1\sum_{ \langle i,j\rangle }\sum_{\sigma} c^{\dagger}_{i\sigma}c_{j\sigma} 
-t_2\sum_{\langle\langle i,j\rangle\rangle}\sum_{\sigma} c^{\dagger}_{i\sigma}c_{j\sigma} \notag \\ 
&+U\sum_{i} n_{i\uparrow}n_{i\downarrow}-\mu\sum_{i,\sigma} n_{i\sigma},
\label{ham}
\end{align} 
where $c^{\dagger}_{i\sigma}$ ($c_{i\sigma}$) creates (annihilates) an electron with spin 
$\sigma$ at site $i$, and $n_{i\sigma}=c^{\dagger}_{i\sigma}c_{i\sigma}$.  
$\langle i,j\rangle$ indicates the nearest-neighbor bonds with the hopping parameter $t_1$ 
and $\langle\langle i,j\rangle\rangle$ indicates the next-nearest-neighbor bonds with the 
hopping parameter $t_2$.  We consider the parameter region $0\leq t_2/t_1\leq 1$, including 
two limiting cases, $t_2=0$ (isotropic triangular lattice) and $t_2=t_1$.  
$U$ is the on-site Coulomb repulsion between two electrons and $\mu$ is the chemical 
potential maintaining the system at half filling.  

In the large-$U$ limit, this model may be mapped onto the triangular-lattice Heisenberg model 
of spin-1/2 defined by the Hamiltonian 
\begin{align}
H=J_1\sum_{\langle i,j\rangle} \bm{S}_{i}\cdot \bm{S}_{j} 
+J_2\sum_{\langle\langle i,j\rangle\rangle} \bm{S}_{i}\cdot \bm{S}_{j}
\label{hei}
\end{align}
with the exchange coupling constants of $J_1=4t^2_1/U$ and $J_2=4t^2_2/U$ for the 
nearest-neighbor and next-nearest-neighbor bonds, respectively.  The spin operator 
is given by $\bm{S}_i=\sum_{\alpha\beta} c^\dagger_{i\alpha}\bm{\sigma}_{\alpha\beta}c_{i\beta}$/2 
with the vector of Pauli matrices $\bm{\sigma}_{\alpha\beta}$.  
The results obtained for the Hubbard model [Eq.~(\ref{ham})] in the strong correlation regime 
are compared with those of the Heisenberg model [Eq.~(\ref{hei})].  

Let us describe the VCA briefly, which is a many-body variational method based on the SFT, 
where the grand potential of the system is formulated as a functional of the self-energy 
\cite{potthoff1,potthoff2,dahnken}.  The ground state of the original system in the thermodynamic 
limit can thus be obtained via the calculation of the grand potential $\Omega$ of the system 
with the exact self-energy.  Then, in the VCA, restricting the trial self-energy to the self-energy 
of the reference system $\Sigma^{\prime}$, we obtain the approximate grand potential as 
\begin{align}
\Omega[\Sigma']=\Omega'+\mathrm{Trln}(G_0^{-1}-\Sigma')^{-1}
-\mathrm{Trln}(G_0^{\prime -1}-\Sigma^{\prime})^{-1}, 
\label{self}
\end{align}
where $\Omega^{\prime}$ is the grand potential of the reference system, and $G_0$ and $G'_0$ 
are the noninteracting Green's functions of the original and reference systems, respectively.  
The Hamiltonian of the reference system $H^{\prime}$ is defined below.  
Note that the short-range correlations within the clusters of the reference system are taken 
into account exactly.  See Refs.~\onlinecite{potthoff3,senechal} for recent reviews of the method.  

The advantage of the VCA is that the spontaneous symmetry breaking can be treated within 
the framework of the theory, where we introduce the Weiss fields as variational parameters.  
In the present case, the Hamiltonian of the reference system is taken as 
$H^{\prime}=H+H_{\rm{M}}$ with the Weiss fields 
\begin{align}
&H_{\mathrm{M}}=H_{\mathrm{120^\circ}}+H_{\mathrm{str}} \\
&H_{\mathrm{120^\circ}}=h'_{\mathrm{120^\circ}}\sum_{i} \bm{e}_{a_i}\cdot \bm{S}_i \\
&H_{\mathrm{str}}=h'_{\mathrm{str}}\sum_{i} e^{i\bm{Q}_{\mathrm{str}}\cdot \bm{r}_i} S^z_i ,
\label{vt}
\end{align}
where $h'_{\mathrm{120^\circ}}$ and $h'_{\mathrm{str}}$ are the strengths of the Weiss fields for 
the 120$^\circ$ N\'eel and stripe ordered states, respectively.   
For the N\'eel order, the unit vectors $\bm{e}_{a_i}$ are rotated by 120$^\circ$ to each other, 
where $a_i$ ($=1,2,3$) is the sublattice index of site $i$.  
For the stripe order, the wave vectors can be taken equivalently as either 
$\bm{Q}_{\rm{str}}=(\pi,\pi/\sqrt{3})$, $(\pi,-\pi/\sqrt{3})$, or $(0,-2\pi/\sqrt{3})$.  
The variational parameters are optimized on the basis of the variational principle, i.e., 
$\partial\Omega/\partial h'=0$, for each magnetic order, where the solution with $h' \ne 0$ 
corresponds to the ordered state.  

In our VCA calculations, we use the 12-site cluster shown in Fig.~\ref{fig2} as the reference 
system.  This is the best appropriate and feasible choice of the reference cluster because we 
can treat the two-sublattice order (stripe order) with an equal number of up- and down-spin 
electrons and the three-sublattice order (120$^\circ$ N\'eel order) with an equal number of 
the three sublattice sites $a_i=1$, 2 and 3.  
The cluster-size and cluster-shape dependences of our results are discussed in Appendix.  
Note that longer period phases such as the spiral phase mentioned in a different system 
\cite{tocchio2} cannot be treated in the present approach; in our analysis, we fix the pitch 
angle of the spiral order to be 120$^\circ$ (or the three-sublattice of $a_i=1,2,3$) even for 
$t_2\neq 0$.  The charge orderings discussed in the extended Hubbard model with 
intersite Coulomb repulsions \cite{tocchio3} are also neglected.  
To our knowledge, no other orders have been predicted in the present triangular-lattice 
Hubbard and Heisenberg models.  

\begin{figure}[thb]
\begin{center}
\includegraphics[width=0.75\columnwidth]{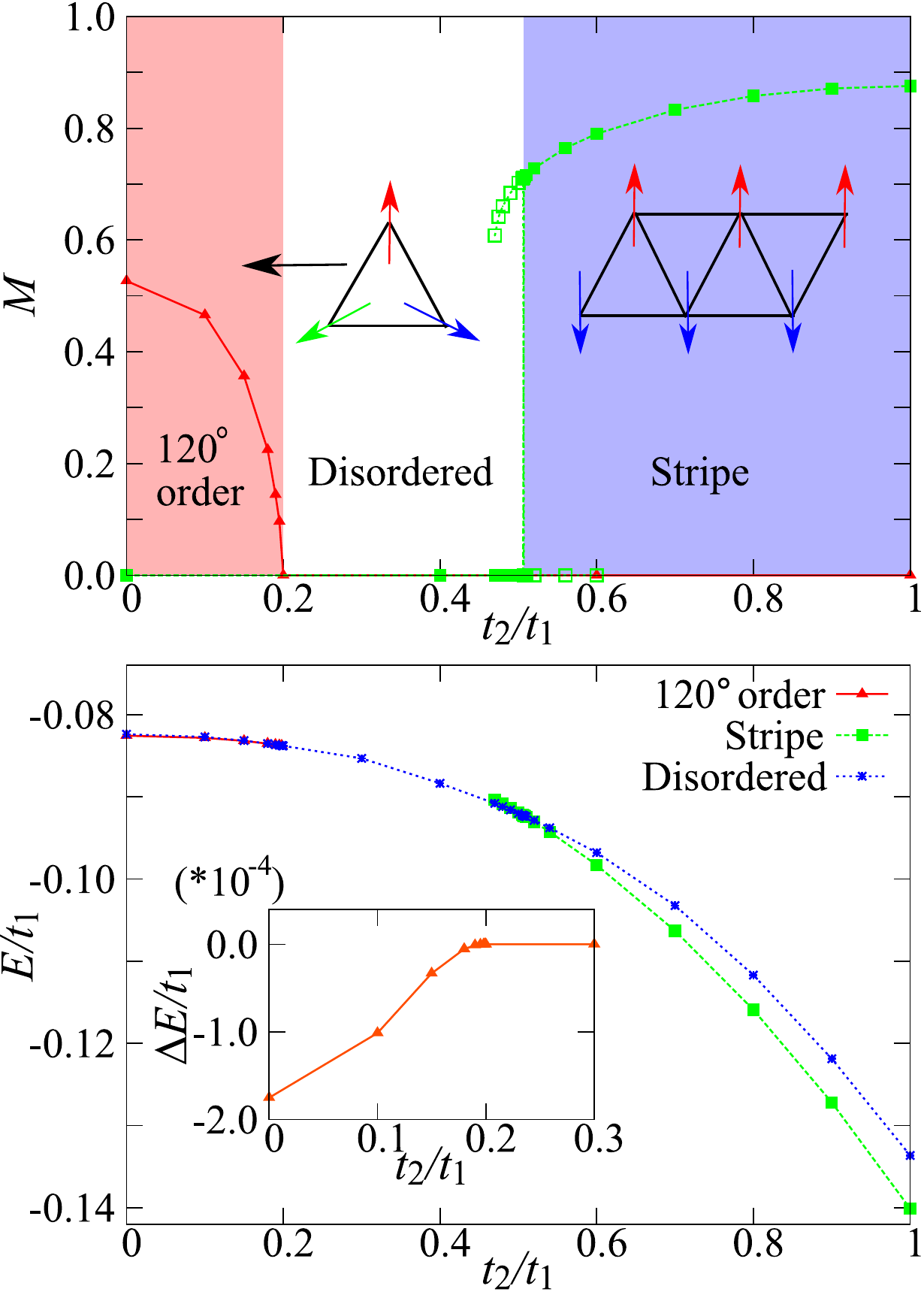}
\caption{(Color online) 
Calculated ground-state phase diagram of our model in the strong correlation regime ($U/t_1=60$).  
Upper panel: the order parameters of the 120$^\circ$ N\'eel and stripe ordered phases.  
Solid (open) symbols indicate that the state is stable (metastable).  
Lower panel: the ground-state energies (per site) of the ordered phases compared with that of 
the disordered phase.  
Inset shows the enlargement of the energy difference $\Delta E$ between the 120$^\circ$ 
ordered and disordered phases.  
}\label{fig3}
\end{center}
\end{figure}

\begin{figure}[thb]
\begin{center}
\includegraphics[width=0.7\columnwidth]{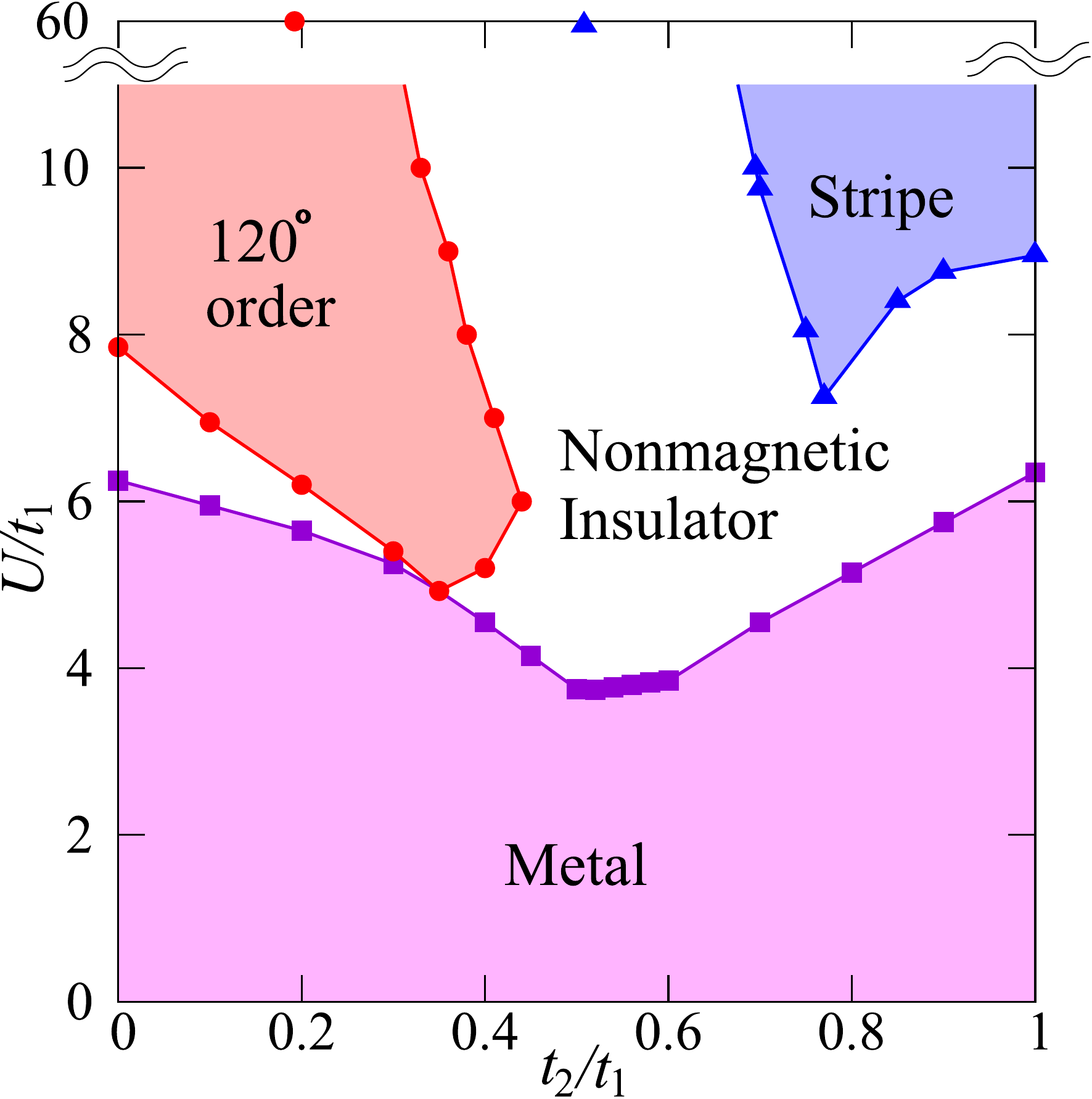}
\caption{(Color online) 
Calculated ground-state phase diagram of our model in the intermediate to weak correlation 
regime, which includes the 120$^\circ$ N\'eel ordered, stripe ordered, nonmagnetic insulating, 
and paramagnetic metallic phases.  Circle and triangle at $U/t_1=60$ indicate the calculated 
phase boundaries of the 120$^\circ$ N\'eel order and stripe order, respectively, shown in 
Fig.~\ref{fig3}.  
}\label{fig4}
\end{center}
\end{figure}

\section{Results of calculation}

\subsection{Strong correlation regime}

First, let us discuss the strong correlation regime, $U/t_1=60$.  We calculate the ground-state 
energies $E=\Omega+\mu$ (per site) and magnetic order parameters $M$ defined as 
$M_{\rm{120^\circ}}=(2/L)\sum_{i} \bm{e}_{a_i}\cdot \langle \bm{S}_i \rangle$ for the 120$^\circ$ N\'eel 
order and $M_{\rm{str}}=(2/L)\sum_{i} e^{i\bm{Q}_{\mathrm{str}}\cdot \bm{r}_i} \langle S^z_i \rangle$ 
for the stripe order, where $\langle\cdots\rangle$ stands for the ground-state expectation 
value.  The results are shown in Fig.~\ref{fig3}, where we find three phases: the 120$^\circ$ 
N\'eel ordered phase around $t_2/t_1=0$, the stripe ordered phase around $t_2/t_1=1$, and 
the nonmagnetic disordered phase in-between.  

At $t_2/t_1=0$, the 120$^\circ$ N\'eel ordered state has the lowest energy and with increasing 
$t_2/t_1$ it approaches the energy of the nonmagnetic disordered state gradually.  
Then, at $t_2/t_1=0.20$, the 120$^\circ$ N\'eel ordered state disappears continuously.  
The calculated order parameter $M_{\rm{120^\circ}}$ also indicates the continuous (or second-order) 
phase transition.  
On the other hand, at $t_2/t_1=1.0$, the stripe ordered state has the lowest energy and, with 
decreasing $t_2/t_1$, the energy of stripe order crosses to that of the nonmagnetic state at 
$t_2/t_1=0.50$, indicating the discontinuous (or first-order) transition between the stripe and 
disordered phases.  The calculated order parameter $M_{\rm{str}}$ also disappears discontinuously 
at $t_2/t_1=0.50$.  

These results may be compared with the previous studies on the $J_1$-$J_2$ triangular-lattice 
Heisenberg model \cite{manuel,mishmash,kaneko,li2,zhu,hu2}.  The transition point between the 
120$^\circ$ N\'eel and nonmagnetic phases has been estimated to be $J_2/J_1=0.05-0.12$, 
which corresponds to $t_2/t_1=0.22-0.35$ of our Hubbard model parameters.  A reasonable 
agreement is thus obtained.  The transition point between the stripe and nonmagnetic phases 
has also been estimated to be $J_2/J_1=0.14-0.19$, which corresponds to $t_2/t_1=0.37-0.44$ 
of our Hubbard model parameters.  We again find a reasonable agreement with our estimation.  
The orders of the phase transitions, i.e., the second-order for the 120$^\circ$ N\'eel phase 
and the first-order for the stripe phase, are also in agreement with the previous study of the 
Heisenberg model \cite{kaneko}.  
We may point out that the strong quantum fluctuations in the frustrated spin degrees of freedom 
causes this nonmagnetic phase because the classical spin model predicts either the 120$^\circ$ 
N\'eel or four-sublattice ordered phase without any intermediate nonmagnetic phases 
\cite{jolicoeur,chubukov,deutscher,lecheminant}.  

\begin{figure}[thb]
\begin{center}
\includegraphics[width=\columnwidth]{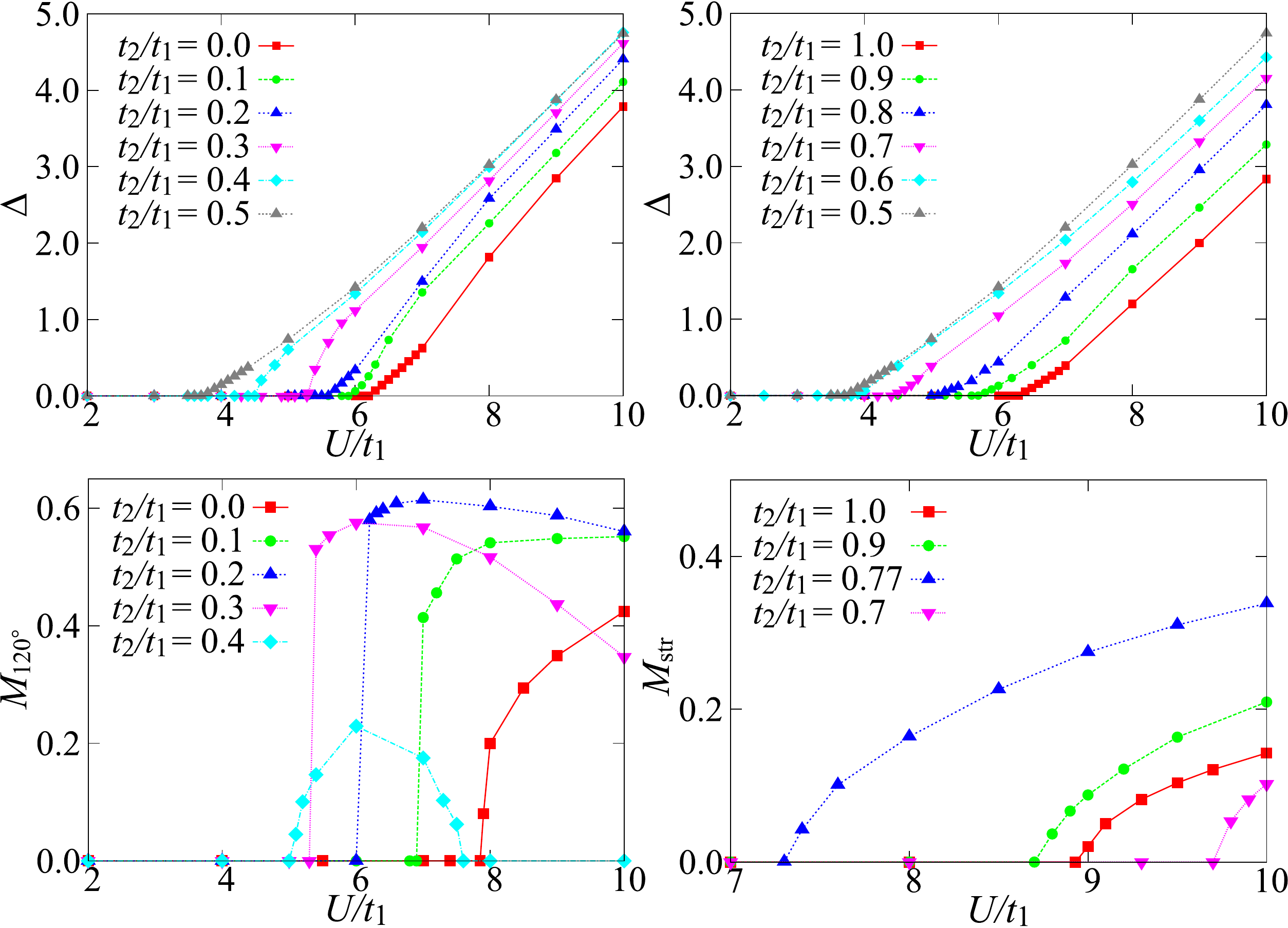}
\caption{(Color online) 
Calculated charge gap $\Delta$ (upper panels) and order parameters $M_{120^\circ}$ and 
$M_\textrm{str}$ (lower panels) of our model as a function of $U/t_1$.  
}\label{fig5}
\end{center}
\end{figure}

\begin{figure}[thb]
\begin{center}
\includegraphics[width=\columnwidth]{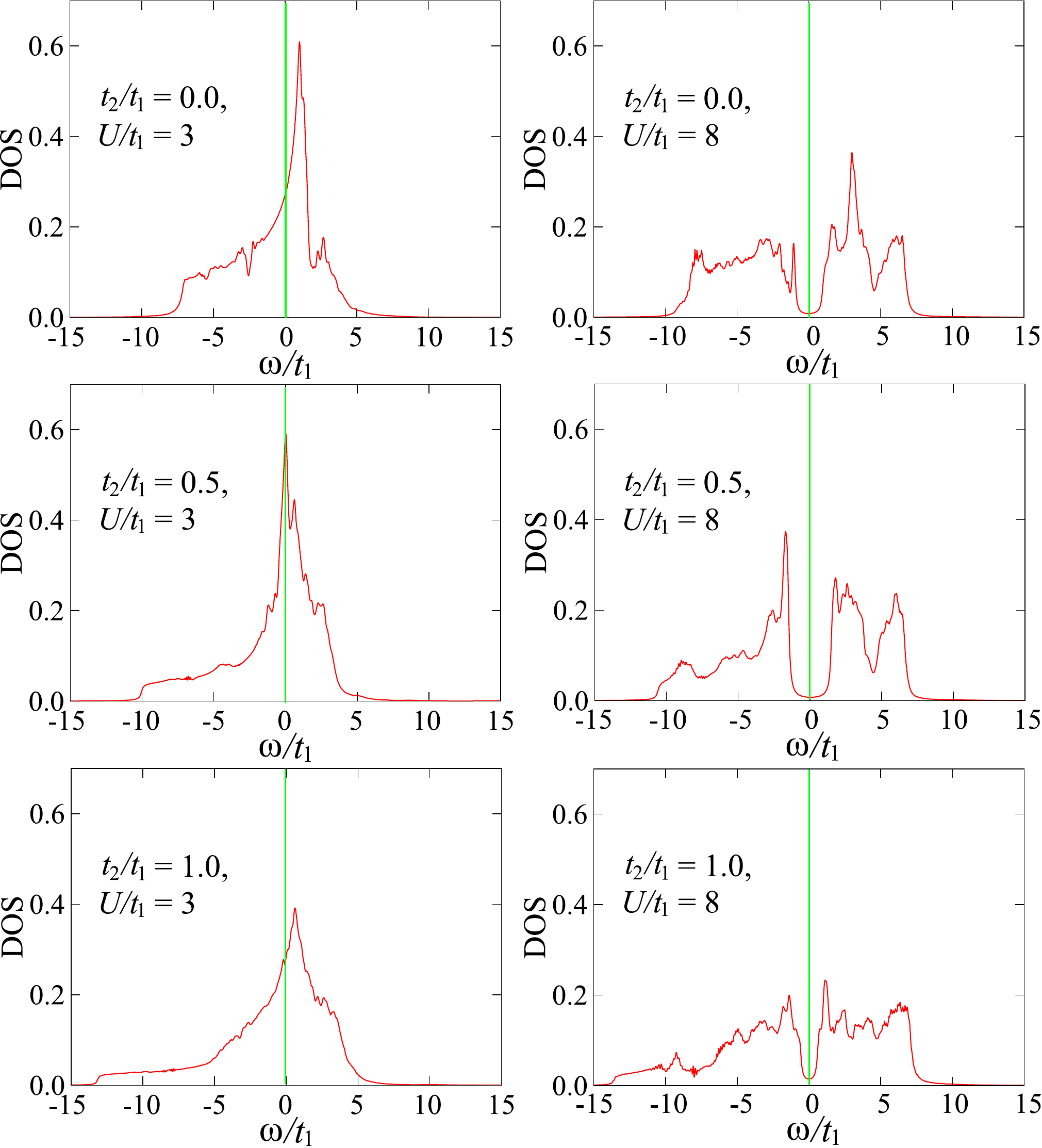}
\caption{(Color online) 
Calculated DOSs of our model in the metallic state (left panels) and insulating state 
without long-range magnetic orders (right panels).  $\eta/t_1=0.1$ is assumed.  
The vertical line in each panel indicates the Fermi level.  
}\label{fig6}
\end{center}
\end{figure}

\begin{figure*}[!t]
\begin{center}
\begin{tabular}{ccc}
\includegraphics[height=0.87\columnwidth]{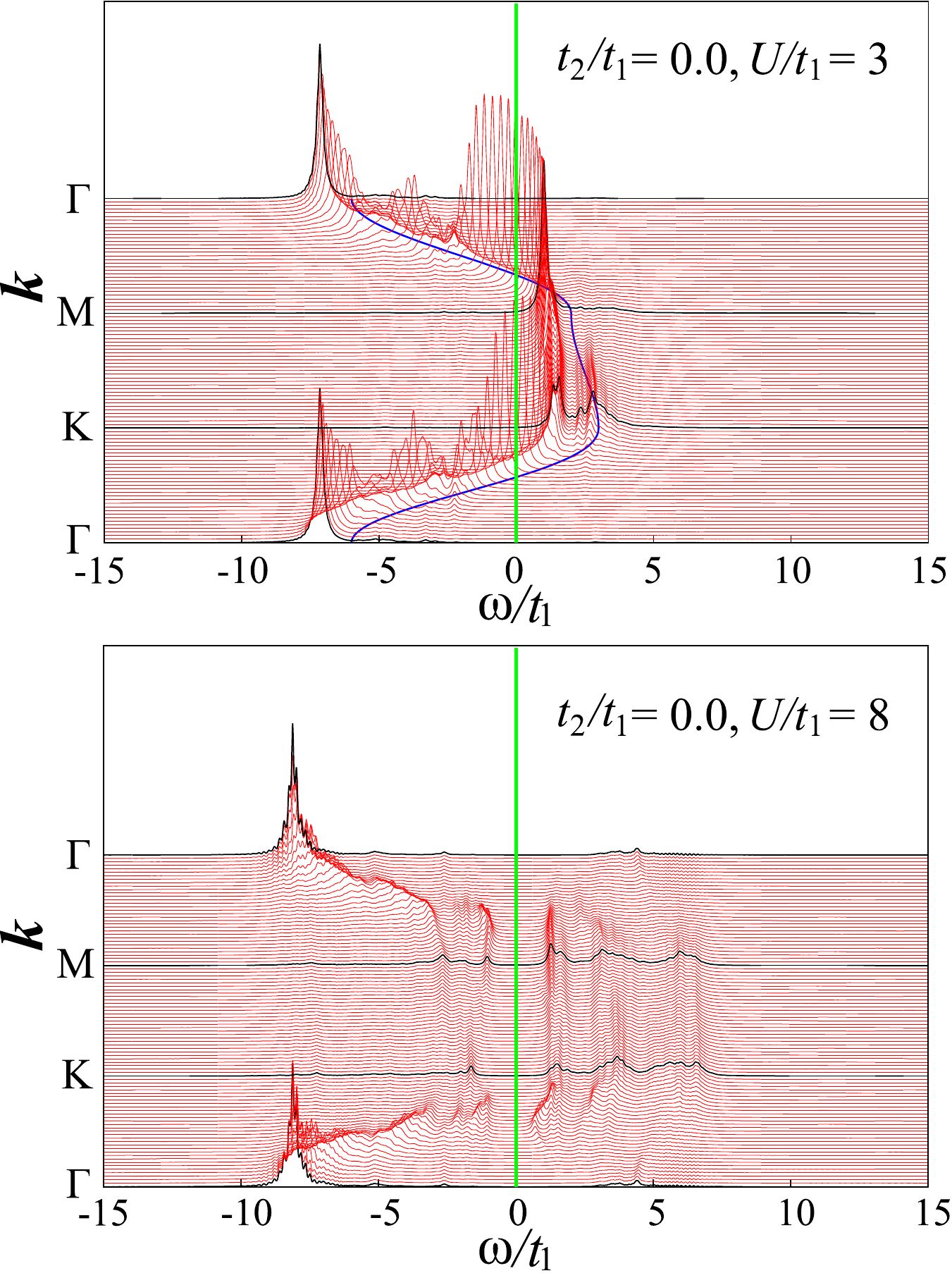}
\includegraphics[height=0.87\columnwidth]{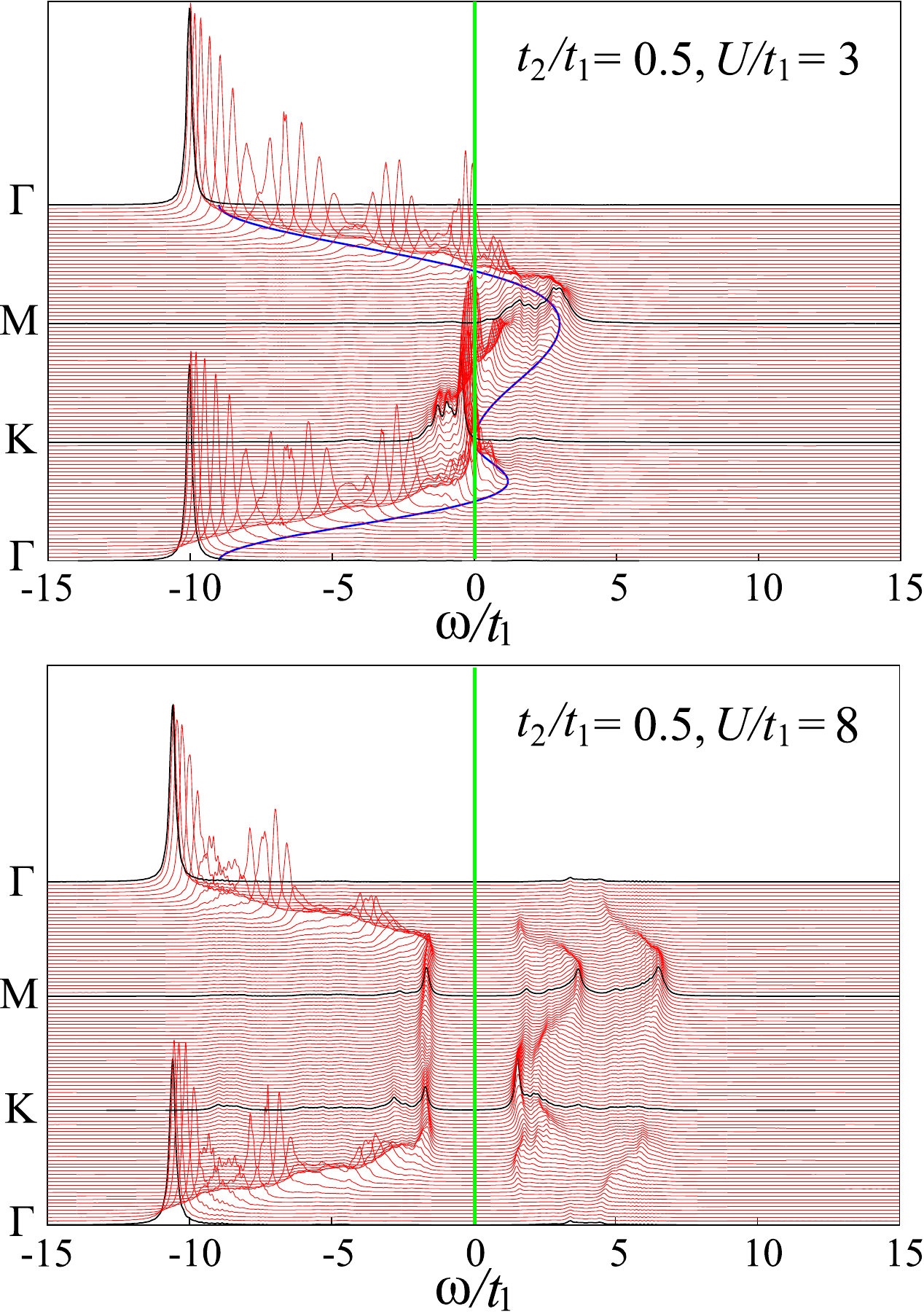}
\includegraphics[height=0.87\columnwidth]{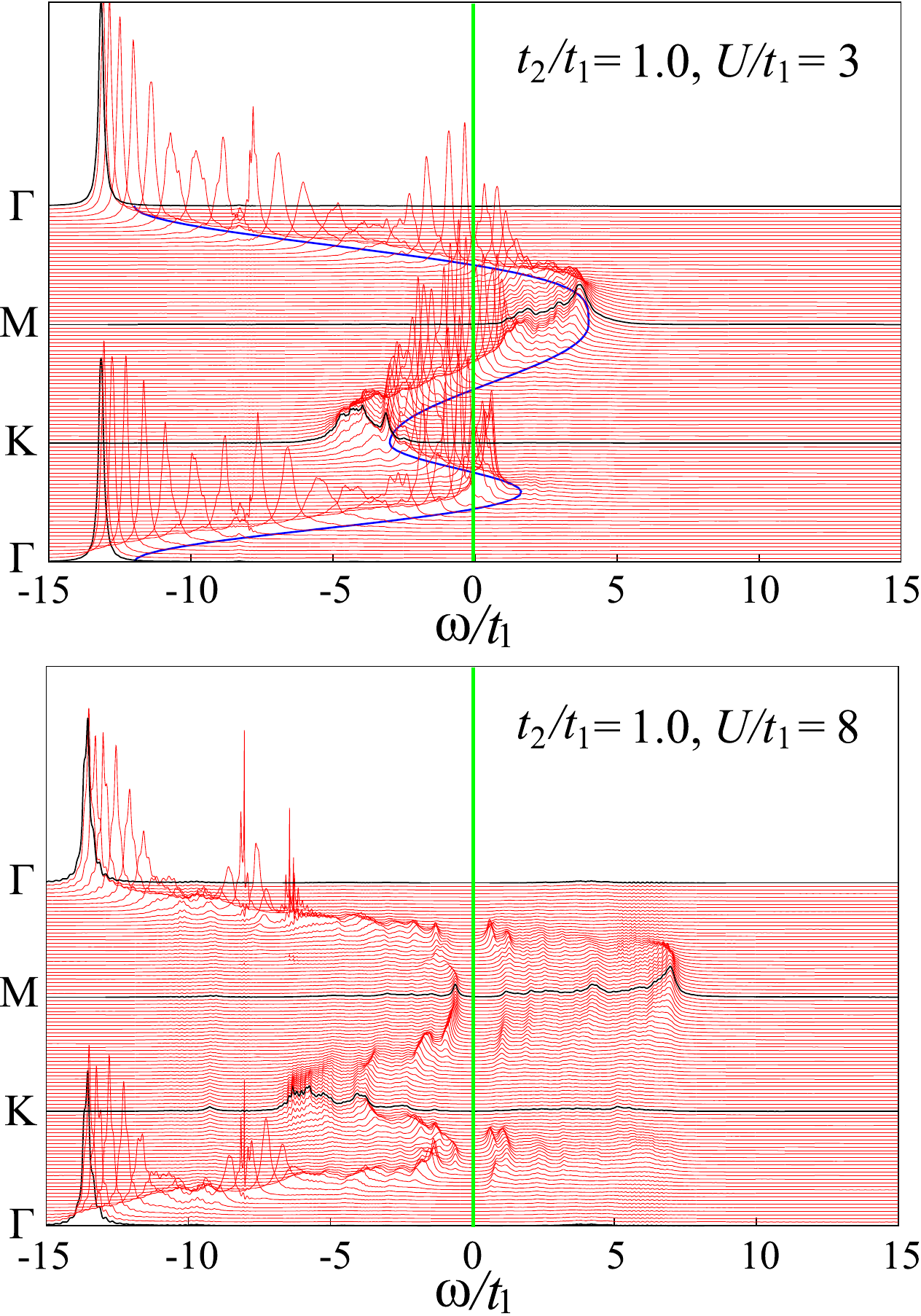}
\end{tabular}
\caption{(Color online) 
Calculated single-particle spectral function $A(\bm{k},\omega)$ in the paramagnetic state 
of our model.  The wave vector $\bm{k}$ is chosen along the line connecting $\Gamma$, K, 
and M points of the Brillouin zone [see Fig.~\ref{fig2} (b)].  $\eta/t_1=0.1$ is assumed.  
The noninteracting band dispersion is also shown by a thin solid curve in each of the upper 
panels.  The Fermi level (indicated by the vertical line) is set at $\omega/t_1=0$.  
}\label{fig7}
\end{center}
\end{figure*}  

\subsection{Intermediate to weak correlation regime}

Next, let us discuss the intermediate to weak correlation regime $0\le U/t_1\le 10$.  
We here calculate the total energies, order parameters, and charge gaps of the model, 
as well as the grand potential as a function of the Weiss fields, and summarize them 
as the ground-state phase diagram in the parameter space $(t_2/t_1, U/t_1)$, as 
shown in Fig.~\ref{fig4}.  
We find four phases: 
the 120$^\circ$ N\'eel and stripe ordered phases at large $U/t_1$, which are 
continuous to the phases at $U/t_1=60$ discussed above, and nonmagnetic insulating 
phase in-between, as well as the paramagnetic metallic phase in the weak correlation 
regime.  
In the intermediate correlation regime, the nonmagnetic insulating phase 
expands to wider parameter regions, which are around $0\le t_2/t_1\alt 0.3$ and 
around $0.4\alt t_2/t_1\le 1$.   
We note that the presence of the 
nonmagnetic insulating phase around $0\le t_2/t_1\alt 0.3$ is in agreement with 
previous studies of the triangular-lattice Hubbard model at $t_2/t_1=0$ 
\cite{sahebsara2,yoshioka1,yoshioka2,yamada1,laubach}.  

The calculated order parameters of the 120$^\circ$ N\'eel and stripe phases are shown 
in Fig.~\ref{fig5} as a function of $U/t_1$ for several values of $t_2/t_1$.  
We find that the transition to the stripe ordered phase is continuous, irrespective of 
$t_2/t_1$, up to a large value of $U/t_1\sim 30$, but it changes to the discontinuous 
transition as seen in Fig.~\ref{fig3} at $U/t_1=60$.  
We also find that the transition to the 120$^\circ$ N\'eel ordered phase is discontinuous 
at $0<t_2/t_1\le 0.35$ for $U/t_1\simeq 6$ but it is continuous for larger values of 
$U/t_1$.  The transition at $U/t_1=60$ is also continuous (see Fig.~\ref{fig3}).  
These behaviors are observed also in the calculated Weiss-field dependence of 
the grand potentials of our model.  

The charge gap is evaluated from the total number of electrons as a function of 
$\mu$ (see Fig.~\ref{fig5}) to examine the Mott metal-insulator transition of the 
system.  We find that the transition is continuous (or second-order) and the phase 
boundary is located around $U/t_1\simeq 4-6$, as shown in Fig.~\ref{fig4}.  
We note that the phase boundary decreases (shifts to a lower $U/t_1$ side) with 
increasing $t_2/t_1$ up to $t_2/t_1\simeq 0.5$, but it increases for larger values of 
$t_2/t_1$.  This behavior is in contrast to that of the square-lattice Hubbard model 
with the next-nearest-neighbor hopping parameters, where a monotonous increase 
in the critical interaction strength is observed \cite{mizusaki,nevidomskyy,yamada2}, 
which is due to the monotonous increase in the band width of the model.  

To find out the origin of this behavior, we calculate the DOS $\rho(\omega)$ and 
single-particle spectral function $A(\bm{k},\omega)$ in the paramagnetic state of 
the system using the CPT, which are defined as 
\begin{align}
&\rho(\omega)=\frac{1}{L}\sum_{\bm{k}}A(\bm{k},\omega)\\
&A(\bm{k},\omega)=-\frac{1}{\pi}\lim_{\eta \rightarrow 0}\Im \mathcal{G}_{\mathrm{CPT}}(\bm{k},\omega+i\eta)
\label{dos}
\end{align}
with the CPT Green's function \cite{senechal} 
\begin{align}
\mathcal{G}_{\mathrm{CPT}}(\bm{k},\omega) = \frac{1}{L_c} \sum_{i,j=1}^{L_c} \mathcal{G}_{ij}(\bm{k},\omega)e^{-i\bm{k}\cdot (\bm{r}_i-\bm{r}_j)} , 
\label{cpt}
\end{align}
where we define the $L_c\times L_c$ matrices for the cluster of size $L_c$ as 
$\mathcal{G}(\bm{k},\omega)=[G'^{-1}(\omega)-V(\bm{k})]^{-1}$ with $V(\bm{k})=G_0^{\prime-1}-G_0^{-1}$.  
The exact Green's function of the reference system $G'(\omega)$ is given by 
\begin{align}
G'_{ij}(\omega)&=\langle\psi_0|c_{i\sigma} \frac{1}{\omega-H'+E_0} c^{\dagger}_{j\sigma}| \psi_0 \rangle \notag \\
&+\langle\psi_0|c^{\dagger}_{j\sigma} \frac{1}{\omega+H'-E_0} c_{i\sigma}| \psi_0 \rangle,
\label{gr}
\end{align}
where $|\psi_0\rangle$ and $E_0$ are the ground state and ground state energy of $H'$. 

The calculated results for the DOS and single-particle spectral function of our model 
are shown in Figs.~\ref{fig6} and \ref{fig7}, respectively.  
We find that the sharp peak appeared above the Fermi level at $t_2/t_1=0$, which is caused by 
the van Hove singularity in the triangular lattice, shifts to the lower energy side with increasing 
$t_2/t_1$, and at $t_2/t_1=0.5$, the peak position coincides with the Fermi level (see Fig.~\ref{fig6}).  
This situation of the high DOS at the Fermi level is energetically unstable \cite{fazekas}, so that 
the band gap opens to gain in the band energy in the presence of the Hubbard interaction $U$.  
With further increasing $t_2/t_1$, the peak shifts to the higher energy side again.  
The Hubbard band gap is then the largest at $t_2/t_1=0.5$ as seen in Figs.~\ref{fig5} and \ref{fig6}.  
This singularity is also seen in the single-particle spectral function as the presence of the 
flat-band region around the K point of the Brillouin zone (see Fig.~\ref{fig7}).  This behavior thus 
explains why the critical interaction strength becomes small at around $t_2/t_1\simeq 0.5$.  

\begin{figure}[thb]
\begin{center}
\includegraphics[width=1.0\columnwidth]{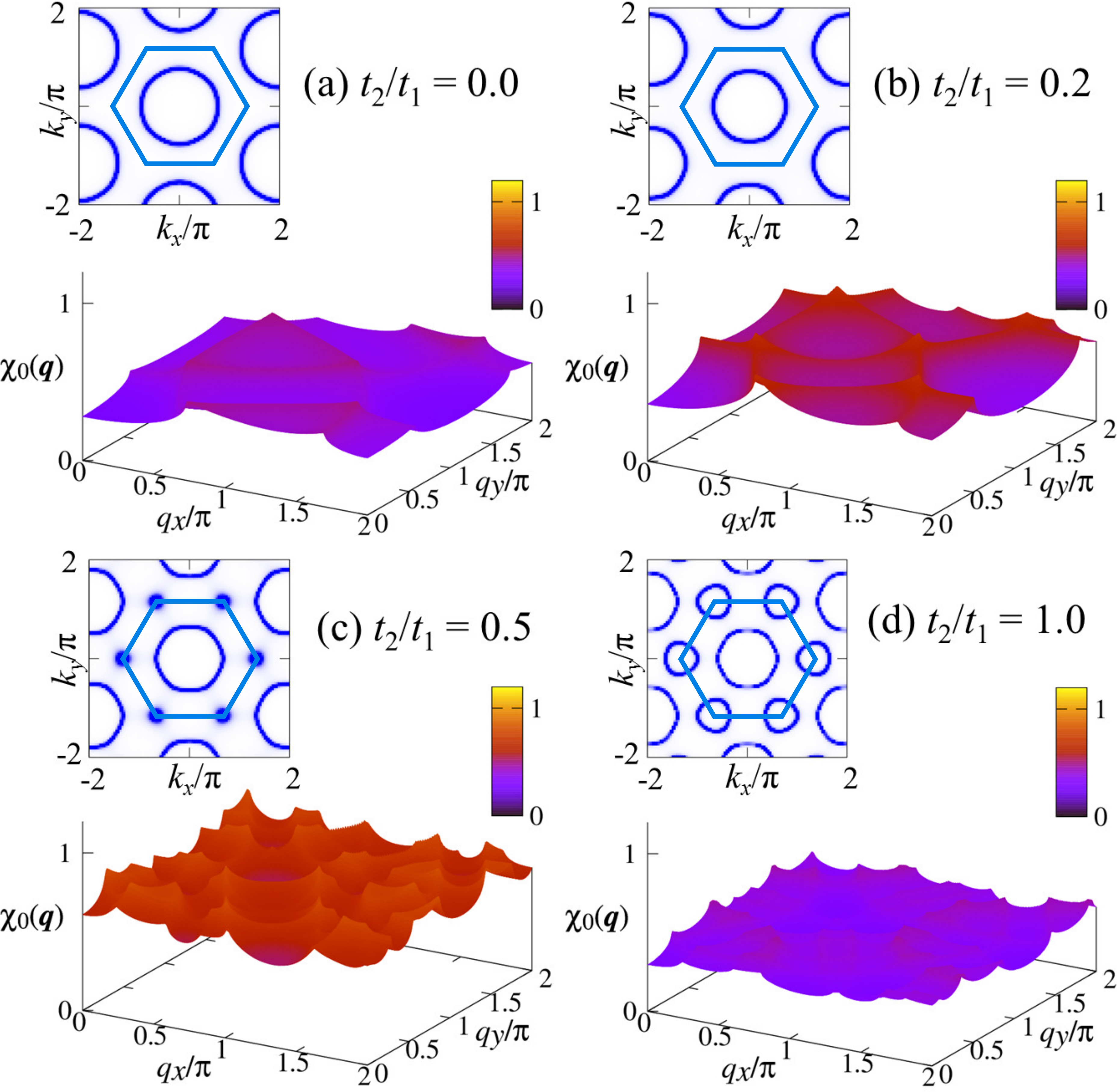}
\caption{(Color online) 
Calculated generalized magnetic susceptibility in the noninteracting limit $\chi_0(\bm{q})$ 
defined in Eq.~(\ref{chiq}).  The corresponding Fermi surface is shown in each panel, where 
the first Brillouin zone is indicated by a hexagon.  
}\label{fig8}
\end{center}
\end{figure}

To confirm the absence of any magnetic instability in our model in the weak correlation regime, 
we here calculate the generalized susceptibility (or Lindhard function) in the noninteracting limit, 
which is defined as 
\begin{align}
\chi_0(\bm{q})=\frac{1}{L}\sum_{\bm{k}} \frac{f(\varepsilon_{\bm{k}})-f(\varepsilon_{\bm{k}+\bm{q}})}
{\varepsilon_{\bm{k}+\bm{q}}-\varepsilon_{\bm{k}}} , 
\label{chiq}
\end{align}
where $\varepsilon_{\bm{k}}$ is the corresponding noninteracting band dispersion and $f(\varepsilon)$ 
is the Fermi function.  The calculated results at temperature $0.01t_1$ are shown in Fig.~\ref{fig8}, 
where we find that, in accordance with the absence of significant Fermi-surface nesting 
features, no singular behaviors actually appear in $\chi_0(\bm{q})$, indicating the absence of 
magnetic long-range orders in the weak correlation limit.  This result supports the validity of 
our phase diagram shown in Fig.~\ref{fig4} in the weak correlation regime.  

\section{Summary}

We have studied the Mott metal-insulator transition and magnetism of the triangular-lattice 
Hubbard model at half filling in the entire region of the interaction strength, taking into account 
the next-nearest-neighbor hopping parameters for the effects of magnetic frustrations.  
We have employed the method of VCA based on the SFT, which has not been used for the 
present purposes.  We have thereby calculated the grand potential of the system as a function 
of the Weiss fields for the 120$^\circ$ N\'eel and stripe magnetic orders, and have determined 
the order parameters.  We have also calculated the DOS and single-particle spectral function as 
well as the charge gap of the system.  These results have been summarized as the ground-state 
phase diagram of the system.  

We have found four phases:
In the strong correlation regime, there appear (i) the 120$^\circ$ N\'eel ordered phase in a 
wide parameter region around $t_2/t_1\simeq 0$ and (ii) the stripe ordered phase in a wide 
parameter region around $t_2/t_1\simeq 1$, and in-between, (iii) the nonmagnetic insulating 
phase caused by the quantum fluctuations in the geometrically frustrated spin degrees of 
freedom emerges.  The obtained phase boundaries in the strong correlation limit have been 
compared with those of the corresponding Heisenberg model to find a reasonable agreement.  
The orders of the phase transitions of the two magnetically ordered phases have also been 
determined.  In the intermediate correlation regime, the nonmagnetic insulating phase expands 
to a wider parameter region of $t_2/t_1$.  
Then, decreasing the interaction strength further, the system turns into (iv) the paramagnetic 
metallic phase in the weak correlation regime via the second-order Mott metal-insulator 
transition.  The characteristic behavior of the critical phase boundary of the Mott transition 
has also been discussed in terms of the shift in the van Hove singularity due to the presence 
of $t_2$, as seen in the calculated DOS and single-particle spectral function.  

We suggest that the phase diagram obtained here may contain different types of nonmagnetic 
insulator (or spin liquid) states depending on the region in the parameter space.  
The characterization of the states is however beyond the scope of the VCA approach 
based on the self-energy (or single-particle Green's function), for which we hope that our 
results will encourage future studies.  

\begin{acknowledgments}
We thank S. Miyakoshi for enlightening discussions and K. Seki for careful reading of our 
manuscript.  This work was supported in part by a Grant-in-Aid for Scientific Research 
(No.~26400349) from JSPS of Japan. 
T.~K.~acknowledges support from the JSPS Research Fellowship for Young Scientists.  
\end{acknowledgments}

\appendix*
\section{Cluster dependence of the phase boundaries}

\begin{figure}[thb]
\begin{center}
\includegraphics[width=0.8\columnwidth]{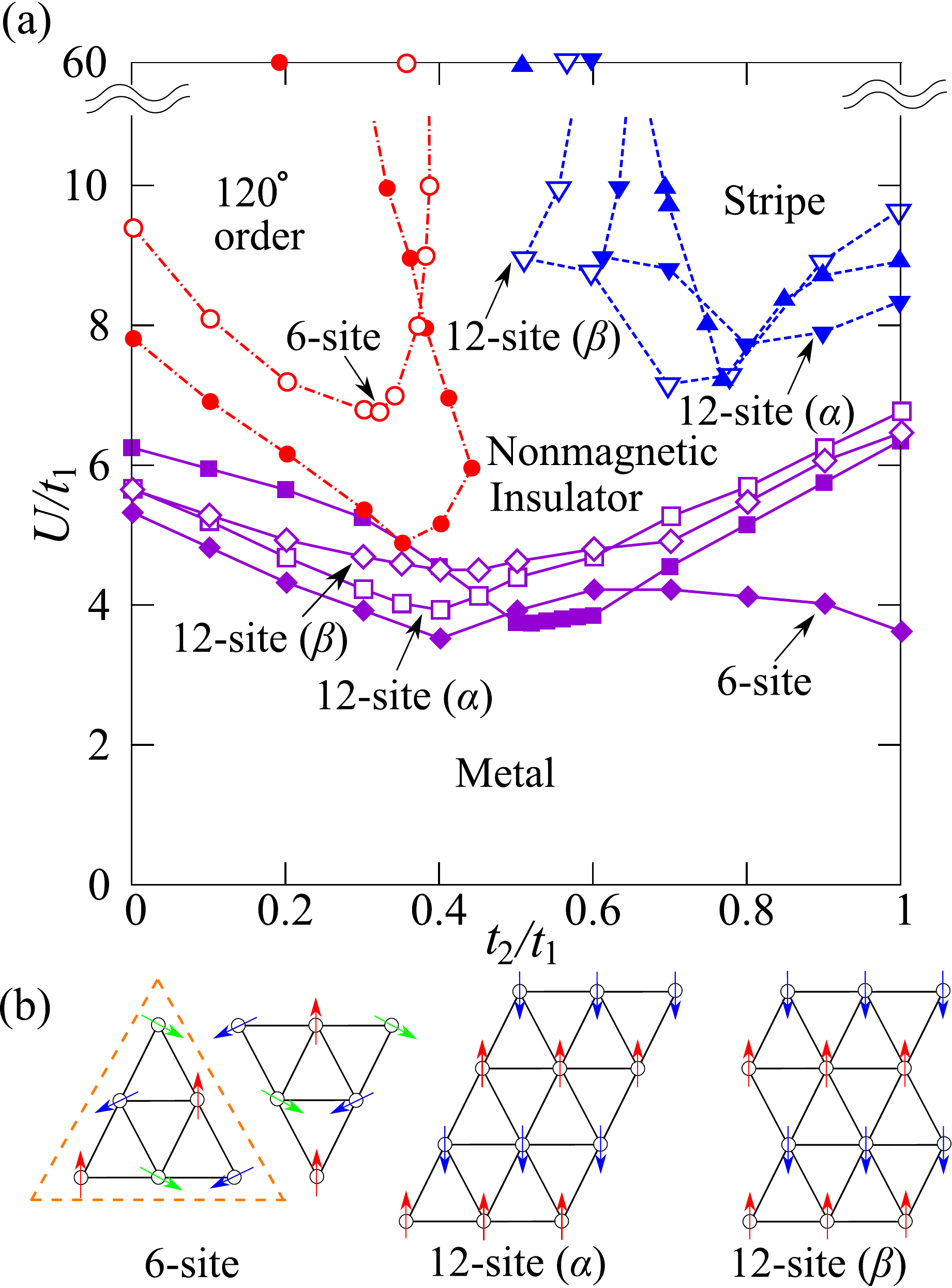}
\caption{(Color online) 
(a) Cluster-size and cluster-shape dependences of the phase boundaries between 
the 120$^\circ$ N\'eel ordered, stripe ordered, nonmagnetic insulating, and 
paramagnetic metallic phases.  Unlabeled lines are the results shown in Fig.~\ref{fig4}.  
(b) Schematic representations of the clusters used in the calculations; in the 
6-site cluster calculation, we combine two of them to reproduce the 
120$^\circ$ N\'eel order.  
}\label{fig9}
\end{center}
\end{figure}

In the VCA, or in any other quantum cluster methods, we can in principle calculate the 
physical quantities in the thermodynamic limit, but the calculated results necessarily 
depend on the size and shape of the solver cluster.  Thus, the choice of the solver 
cluster is important in the present approach.  
In the main text, we have chosen the 12-site cluster shown in Fig.~\ref{fig2}, which is the 
best appropriate one because it is first of all computationally feasible and also because it fits 
with both the three-sublattice 120$^\circ$ order and two-sublattice stripe order without 
introducing unnecessary frustrations.  
However, it seems instructive to check the cluster-size and cluster-shape dependences 
of our results presented in the main text.  
\\\indent
Here, we choose several clusters [see Fig.~\ref{fig9}(b)] that fit either with the 120$^\circ$ 
order or with the stripe order, and we calculate the phase boundaries to check the solver 
cluster dependence of the ground-state phase diagram shown in Fig.~\ref{fig4}.  
The calculated results for the phase boundaries are shown in Fig.~\ref{fig9}(a).  
We thus find that the phase boundary between the 120$^\circ$ N\'eel ordered and 
nonmagnetic insulating phases is located around $0\le t_2/t_1\alt 0.4$ in an intermediate 
to large $U/t_1$ $(\agt 6)$ region and that the phase boundary between the stripe ordered 
and nonmagnetic insulating phases is located around $0.5\alt t_2/t_1\le 1$ in an intermediate 
to large $U/t_1$ $(\agt 7)$ region, irrespective of the appropriate choices of the solver cluster.  
We also find that the phase boundary of the Mott metal-insulator transition is located 
around $U/t_1\simeq 4-6$ with a minimum at $t_2/t_1\simeq 0.5$, irrespective of the 
choices of the solver cluster.


\begin{thebibliography}{99}
%
\bibitem{anderson} P. Anderson, Mater. Res. Bull. \textbf{8}, 153 (1973).
%
\bibitem{balents} L. Balents, Nature (London) \textbf{464}, 199 (2010).
%
\bibitem{lee} P. A. Lee, N. Nagaosa, and X. -G. Wen, Rev. Mod. Phys. \textbf{78}, 17 (2006).
%
\bibitem{mott} N. Mott, \textit{Metal-Insulator Transitions} (Taylor and Francis, London, UK, 1990).  
%
\bibitem{imada} M. Imada, A. Fujimori, and Y. Tokura, Rev. Mod. Phys. \textbf{70}, 1039 (1998).  
%
\bibitem{morita} H. Morita, S. Watanabe, and M. Imada, J. Phys. Soc. Jpn. \textbf{71}, 2109 (2002).  
%
\bibitem{sahebsara} P. Sahebsara and D. S\'en\'echal, Phys. Rev. Lett. \textbf{97}, 257004 (2006). 
%
\bibitem{watanabe} T. Watanabe, H. Yokoyama, Y. Tanaka, and J. Inoue, 
Phys. Rev. B \textbf{77}, 214505 (2008).
%
\bibitem{ohashi} T. Ohashi, T. Momoi, H. Tsunetsugu, and N. Kawakami, Phys. Rev. Lett. \textbf{100}, 076402 (2008). 
%
\bibitem{tocchio1} L. F. Tocchio, A. Parola, C. Gros, and F. Becca,  
Phys. Rev. B \textbf{80}, 064419 (2009).
%
\bibitem{tocchio2} L. F. Tocchio, H. Feldner, F. Becca, R. Valenti, and C. Gros, 
Phys. Rev. B \textbf{87}, 035143 (2013).
%
\bibitem{yamada1} A. Yamada, Phys. Rev. B \textbf{89}, 195108 (2014).
%
\bibitem{laubach} M. Laubach, R. Thomale, C. Platt, W. Hanke, and G. Li,  
Phys. Rev. B \textbf{91}, 245125 (2015).  
%
\bibitem{misumi} K. Misumi, T. Kaneko, and Y. Ohta, J. Phys. Soc. Jpn., in press (2016).  
%
\bibitem{komatsu} T. Komatsu, N. Matsukawa, T. Inoue, and G. Saito,
J. Phys. Soc. Jpn. \textbf{65}, 1340 (1996).
%
\bibitem{shimizu} Y. Shimizu, K. Miyagawa, K. Kanoda, M. Maesato, 
and G. Saito, Phys. Rev. Lett. \textbf{91}, 107001 (2003).
%
\bibitem{kurosaki} Y. Kurosaki, Y. Shimizu, K. Miyagawa, K. Kanoda, 
and G. Saito, Phys. Rev. Lett. \textbf{95}, 177001 (2005).
%
\bibitem{manna} R. S. Manna, M. de Souza, A. Br\"uhl, J. A. Schlueter, 
and M. Lang, Phys. Rev. Lett. \textbf{104}, 016403 (2010).
%
\bibitem{itou} T. Itou, A. Oyamada, S. Maegawa, M. Tamura, and R. Kato, 
Phys. Rev. B \textbf{77}, 104413 (2008).
%
\bibitem{yamashita} M. Yamashita, N. Nakata, Y. Senshu, M. Nagata, 
H. M. Yamamoto, R. Kato, T. Shibauchi, and Y. Matsuda, 
Science \textbf{328}, 1246 (2010).
%
\bibitem{weihong} Z. Weihong, R. H. McKenzie, and R. R. P. Singh, Phys. Rev. B \textbf{59}, 14367 (1999).
%
\bibitem{hauke1} P. Hauke, T. Roscilde, V. Murg, J. Cirac, and R. Schmied, 
New J. Phys. \textbf{13}, 075017 (2011).
%
\bibitem{hauke2} P. Hauke, Phys. Rev. B \textbf{87}, 014415 (2013).
%
\bibitem{ma} A rare exception is Ba$_3$CoSb$_2$O$_9$, which was reported to be an almost perfect 
realization of a spin-1/2 equilateral triangular-lattice antiferromagnet. 
See, e.g., J. Ma, Y. Kamiya, T. Hong, H. B. Cao, G. Ehlers, W. Tian, C. D. Batista, Z. L. Dun, 
H. D. Zhou, and M. Matsuda, Phys. Rev. Lett. \textbf{116}, 087201 (2016).  
%
\bibitem{jolicoeur} T. Jolicoeur, E. Dagotto, E. Gagliano, and S. Bacci,  
Phys. Rev. B \textbf{42}, 4800 (1990).
%
\bibitem{chubukov} A. V. Chubukov and T. Jolicoeur, Phys. Rev. B \textbf{46}, 11137 (1992).
%
\bibitem{deutscher} R. Deutscher and H. U. Everts, Z. Phys. B \textbf{93}, 77 (1993).
%
\bibitem{lecheminant} P. Lecheminant, B. Bernu, C. Lhuillier, and L. Pierre,  
Phys. Rev. B \textbf{52}, 6647 (1995).
%
\bibitem{manuel} L. O. Manuel and H. A. Ceccatto, Phys. Rev. B \textbf{60}, 9489 (1999).
%
\bibitem{mishmash} R. V. Mishmash, J. R. Garrison, S. Bieri, and C. Xu,
Phys. Rev. Lett. \textbf{111}, 157203 (2013).
%
\bibitem{kaneko} R. Kaneko, S. Morita, and M. Imada,
J. Phys. Soc. Jpn. \textbf{83}, 093707 (2014).
%
\bibitem{li2} P. H. Y. Li, R. F. Bishop, and C. E. Campbell, 
Phys. Rev. B \textbf{91}, 014426 (2015).
%
\bibitem{zhu} Z. Zhu and S. R. White,  
Phys. Rev. B \textbf{92}, 041105 (2015).
%
\bibitem{hu2} W.-J. Hu, S.-S. Gong, W. Zhu, and D. N. Sheng,  
Phys. Rev. B \textbf{92}, 140403 (2015). 
%
\bibitem{iqbal} Y. Iqbal, W.-J. Hu, R. Thomale, D. Poilblanc, and F. Becca, Phys. Rev. B \textbf{93}, 144411 (2016).  
%
\bibitem{potthoff1} M. Potthoff, M. Aichhorn, and C. Dahnken, 
Phys. Rev. Lett. \textbf{91}, 206402 (2003).  
%
\bibitem{potthoff2} M. Potthoff, Eur. Phys. J. B \textbf{32}, 429 and \textbf{36} 335 (2003).
%
\bibitem{dahnken} C. Dahnken, M. Aichhorn, W. Hanke, E. Arrigoni, and M. Potthoff, 
Phys. Rev. B \textbf{70}, 245110 (2004).  
%
\bibitem{potthoff3} M. Potthoff, in \textit{Strongly Correlated Systems--Theoretical Methods}, 
Vol.~171 of Springer Series in Solid-State Sciences (Springer-Verlag, Berlin Heidelberg, 2012), Chap.~10, pp.~303--339.  
%
\bibitem{senechal} D. S\'en\'echal, in \textit{Strongly Correlated Systems--Theoretical Methods}, 
Vol.~171 of Springer Series in Solid-State Sciences (Springer-Verlag, Berlin Heidelberg, 2012), Chap.~8, pp.~237--270.  
%
%
%
\bibitem{tocchio3} L. F. Tocchio, C. Gros, X.-F. Zhang, and S. Eggert, Phys. Rev. Lett. \textbf{113}, 246405 (2014). 
%
\bibitem{sahebsara2} P. Sahebsara and D. S\'en\'echal, Phys. Rev. Lett. \textbf{100}, 136402 (2008). 
%
\bibitem{yoshioka1} T. Yoshioka, A. Koga, and N. Kawakami, Phys. Rev. Lett. \textbf{103}, 036401 (2009). 
%
\bibitem{yoshioka2} T. Yoshioka, A. Koga, and N. Kawakami, Phys. Stat. Sol. (b) \textbf{247}, 635 (2010).  
%
\bibitem{mizusaki} T. Mizusaki and M. Imada,
Phys. Rev. B \textbf{74}, 014421 (2006).
%
\bibitem{nevidomskyy} A. H. Nevidomskyy, C. Scheiber, D. S\'en\'echal, and A. -M. S. Tremblay, 
Phys. Rev. B \textbf{77}, 064427 (2008).
%
\bibitem{yamada2} A. Yamada, K. Seki, R. Eder, and Y. Ohta, 
Phys. Rev. B \textbf{88}, 075114 (2013).
%
\bibitem{fazekas} P. Fazekas, \textit{Lecture Notes on Electron Correlation and Magnetism}, 
Vol.~5 of Series in Modern Condensed Matter Physics (World Scientific, Singapore, 1999).  
%
\end{thebibliography}
\end{document}